# Generalized Multiple Access Channels with Confidential Messages

Yingbin Liang and H. Vincent Poor
Department of Electrical Engineering
Princeton University
Princeton, NJ 08536, USA
Email: {yingbinl,poor}@princeton.edu

*Abstract*— **A discrete memoryless generalized multiple access channel (GMAC) with confidential messages is studied, where two users attempt to transmit common information to a destination and each user also has private (confidential) information intended for the destination. This channel generalizes the multiple access channel (MAC) in that the two users also receive channel outputs. It is assumed that each user views the other user as a wire-tapper, and wishes to keep its confidential information as secret as possible from the other user. The level of secrecy of the confidential information is measured by the equivocation rate. The performance measure of interest is the rate-equivocation tuple that includes the common rate, two private rates and two equivocation rates as components. The set that includes all achievable rate-equivocation tuples is referred to as the capacity-equivocation region. For the GMAC with one confidential message set, where only one user (user 1) has private (confidential) information for the destination, inner and outer bounds on the capacity-equivocation region are derived. The outer bound provides a tight converse to the secrecy capacity region, which is the set of all achievable rates with user 2 being perfectly ignorant of confidential messages of user 1, thus establishing the secrecy capacity region. Furthermore, the degraded GMAC with one confidential message set is further studied, and the capacity-equivocation region and the secrecy capacity region are established. For the GMAC with two confidential message sets, where both users have confidential messages for the destination, an inner bound on the capacity-equivocation region is obtained. The secrecy rate region is derived, where each user's confidential information is perfectly hidden from the other user.**

## I. Introduction

Two important issues in communications are reliability and security. The reliability quantifies the maximum achievable rate (capacity) with small probability of error, and has been studied intensively since Shannon theory was established [1]. Security is an important issue when the transmitted information is confidential and needs to be kept as secret as possible from wire-tappers or eavesdroppers. The level of secrecy of confidential information at a wire-tapper can be measured by the equivocation rate, i.e., the entropy rate of confidential messages conditioned on channel outputs at the wire-tapper. If both reliability and security are considered, the performance measure of interest is the rate-equivocation tuple that includes both the communication rates and the equivocation rates (achieved at wire-tappers) as components. We refer to the set

This research was supported by the National Science Foundation under Grant ANI-03-38807.

that consists of all achievable rate-equivocation tuples as the *capacity-equivocation region*.

Communication of confidential messages has been studied in the literature for some classes of channels. The wire-tap channel was introduced by Wyner in [2], where a sender wishes to transmit information to a legitimate receiver and to keep a wire-tapper as ignorant of this information as possible. The channel from the sender to the legitimate receiver and the wire-tapper was assumed to be a degraded broadcast channel. The trade-off between the communication rate to the legitimate receiver and the level of ignorance at the wire-tapper was developed.

The broadcast channel with confidential messages was studied in [3] as a generalization of the wire-tap channel, where the sender also wishes to transmit common information to both the legitimate receiver and the wire-tapper in addition to the private (confidential) information to the legitimate receiver. Moreover, the broadcast channel from the sender to the two receivers was assumed to be general and may not be degraded. The capacity-equivocation region was established for this channel, and the secrecy capacity region was given. The relay channel with confidential messages was studied in [4], where the relay node acts as both a helper and a wire-tapper.

In this paper, we consider a two-user *generalized multiple access channel (GMAC) with confidential messages*, which generalizes the multiple access channel (MAC) [5, Sec. 14.3] by allowing both users to receive noisy channel outputs. This channel model is motivated by wireless communications, where transmitted signals are broadcast over open media and can be received by all nodes within communication range. For this channel, we assume that users 1 and 2 have common information and each user has its private (confidential) information intended for a destination. Since two users also receive channel outputs, they may extract each other's confidential information from their received channel outputs. However, each user treats the other user as a wire-tapper, and wishes to keep this wire-tapper as ignorant of its confidential message as possible. The level of ignorance of one user's confidential message at the other user (wire-tapper) is measured by the equivocation rate. Our goal is to study the capacity-equivocation region of the GMAC with confidential messages.

We first study the GMAC with one confidential message set, where two users have common information for the destination and only one user (user 1) has private (confidential) information for the destination. For this case, we obtain inner and outer bounds on the capacity-equivocation region. The two bounds match partially and determine the capacity-equivocation region partially. Furthermore, the outer bound provides a tight converse to the secrecy capacity region, which is the set of all achievable rates with user 2 being perfectly ignorant of confidential messages of user 1, and we hence establish the secrecy capacity region. We also study the degraded GMAC with one confidential message set, where outputs at user 2 are degraded versions of outputs at the destination. For this channel, we show a tight converse and establish the capacity-equivocation region and the secrecy capacity region.

We further study the general case of the GMAC with two confidential message sets, where both users have confidential messages for the destination in addition to common messages. We obtain an achievable rate-equivocation region (inner bound on the capacity-equivocation region). We demonstrate a trade-off between the two equivocation rates corresponding to the two sets of confidential messages sent by user 1 and user 2. Based on the rate-equivocation region, we derive the secrecy rate region, where confidential messages of each user are perfectly secret from the other user.

In this paper, we use $x^n$ to indicate the vector $(x_1, \ldots, x_n)$, and use $x_i^n$ to indicate the vector $(x_i, \ldots, x_n)$.

The organization of this paper is as follows. In Section II, we introduce the channel model of the GMAC with confidential messages. In Section III, we present our results for the GMAC with one confidential message set. In Section IV, we present our results for the degraded GMAC with one confidential message set. In Section V, we present our results for the general case of the GMAC with two confidential message sets. In the final section, we give concluding remarks. We omit the proofs due to space limitations. The details can be found in [6].

## II. CHANNEL MODEL

In this section, we first define the GMAC, and then define the performance measure of interest for the GMAC with confidential messages.

*Definition 1:* A discrete memoryless GMAC consists of two finite channel input alphabets $\mathcal{X}_1$ and $\mathcal{X}_2$, three finite channel output alphabets $\mathcal{Y}, \mathcal{Y}_1$ and $\mathcal{Y}_2$, and a transition probability distribution $p(y, y_1, y_2 | x_1, x_2)$ (see Fig. 1), where $x_1 \in \mathcal{X}_1$ and $x_2 \in \mathcal{X}_2$ are channel inputs from users 1 and 2, respectively, and $y \in \mathcal{Y}, y_1 \in \mathcal{Y}_1$ and $y_2 \in \mathcal{Y}_2$ are channel outputs at the destination, user 1 and user 2, respectively.

*Definition 2:* A $(2^{nR_0}, 2^{nR_1}, 2^{nR_2}, n)$ code for the GMAC consists of the following:
- Three message sets: $\mathcal{W}_0 = \{1, 2, \ldots, 2^{nR_0}\}$, $\mathcal{W}_1 = \{1, 2, \ldots, 2^{nR_1}\}$ and $\mathcal{W}_2 = \{1, 2, \ldots, 2^{nR_2}\}$. The common message $W_0$ and private messages $W_1$ and $W_2$ are independent and uniformly distributed over the message sets $\mathcal{W}_0, \mathcal{W}_1$ and $\mathcal{W}_2$, respectively.
- Two (stochastic) encoders, one at user 1: $\mathcal{W}_0 \times \mathcal{W}_1 \to \mathcal{X}_1^n$, which maps each message pair $(w_0, w_1) \in \mathcal{W}_0 \times \mathcal{W}_1$ to a codeword $x_1^n \in \mathcal{X}_1^n$; the other at user 2: $\mathcal{W}_0 \times \mathcal{W}_2 \to \mathcal{X}_2^n$, which maps each message pair $(w_0, w_2) \in \mathcal{W}_0 \times \mathcal{W}_2$ to a codeword $x_2^n \in \mathcal{X}_2^n$;
- One decoder at the destination: $\mathcal{Y}^n \to \mathcal{W}_0 \times \mathcal{W}_1 \times \mathcal{W}_2$, which maps a received sequence $y^n$ to a message tuple $(w_0, w_1, w_2) \in \mathcal{W}_0 \times \mathcal{W}_1 \times \mathcal{W}_2$.

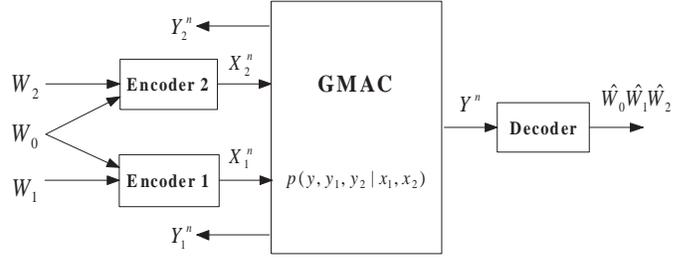

Fig. 1. Generalized multiple access channel

Note that in the GMAC although users 1 and 2 can receive channel outputs (see Fig. 1), they are only passive listeners in that their encoding functions are not affected by these received outputs. However, since outputs at each user contain the other user's private (confidential) information, each user may extract the other user's private (confidential) information from its outputs. We assume that each user treats the other user as a wire-tapper, and wishes to keep the other user as ignorant of its private (confidential) messages as possible. We hence define the following two equivocation rates:

$$\begin{aligned} \text{at user 2: } & \frac{1}{n} H(W_1 | Y_2^n, X_2^n, W_0, W_2) \\ \text{at user 1: } & \frac{1}{n} H(W_2 | Y_1^n, X_1^n, W_0, W_1) \end{aligned} \quad (1)$$

which indicate the level of ignorance of the confidential message $W_1$ at user 2 and the level of ignorance of the confidential message $W_2$ at user 1, respectively. Note that the larger the equivocation rate, the higher the level of secrecy.

For the GMAC with confidential messages, a rate-equivocation tuple $(R_0, R_1, R_2, R_{1,e}, R_{2,e})$ is *achievable* if there exists a sequence of $(2^{nR_0}, 2^{nR_1}, 2^{nR_2}, n)$ codes with the average error probability $P_e^{(n)} \to 0$ as $n$ goes to infinity and with the equivocation rates $R_{1,e}$ and $R_{2,e}$ satisfying

$$\begin{aligned} \lim_{n \to \infty} \frac{1}{n} H(W_1 | Y_2^n, X_2^n, W_0, W_2) &\geq R_{1,e} \\ \lim_{n \to \infty} \frac{1}{n} H(W_2 | Y_1^n, X_1^n, W_0, W_1) &\geq R_{2,e}. \end{aligned} \quad (2)$$

Note that the rate-equivocation tuple $(R_0, R_1, R_2, R_{1,e}, R_{2,e})$ includes both the reliable communication rates and the equivocation rates, and it indicates the common and private rates $(R_0, R_1, R_2)$ achieved at certain levels of communication secrecy $(R_{1,e}, R_{2,e})$.

The capacity-equivocation region, denoted by $\mathscr{C}$, is the closure of the set that consists of all achievable rate-equivocation tuples $(R_0, R_1, R_2, R_{1,e}, R_{2,e})$. Our goal is to study the capacity-equivocation region of the GMAC with confidential messages.

## III. GMAC WITH ONE CONFIDENTIAL MESSAGE SET

In this section, we study the GMAC with one confidential message set, where user 2 has only common messages and does not have confidential messages for the destination. This model generalizes the MAC with degraded message sets studied in [7] to consider the private messages sent from user 1 to be confidential, i.e., needing to be as secret as possible from user 2. This model is also a counterpart of the broadcast channel with confidential messages studied in [3].

For the GMAC with one confidential message set, the rate $R_2 = 0$, and the equivocation rate $R_{2,e}$ is not of interest. Hence channel outputs at user 1 do not play roles in the analysis. For notational convenience, we use $R_e$ to indicate $R_{1,e}$ in this case. Now the rate-equivocation tuple becomes $(R_0, R_1, R_e)$; i.e., it contains three components. The following two theorems provide inner and outer bounds on the capacity-equivocation region $\mathscr{C}^I$ of the GMAC in this situation.

*Theorem 1:* The following convexified region is an inner bound on the capacity-equivocation region $\mathscr{C}^I$ for the GMAC with one confidential message set:

$$\mathscr{R}^I = \textbf{Convex} \bigcup_{p(q,x_2)p(u|q)p(x_1|u)p(y,y_2|x_1,x_2)}$$
$$\left\{\begin{array}{l} (R_0, R_1, R_e): \\ R_0 \geq 0, \ R_1 \geq 0, \\ R_1 \leq I(U;Y|X_2,Q), \\ R_0 + R_1 \leq I(U,X_2,Q;Y), \\ 0 \leq R_e \leq R_1, \\ R_e \leq [I(U;Y|X_2,Q) - I(U;Y_2|X_2,Q)]_+, \\ R_e \leq [I(U,X_2,Q;Y) - R_0 - I(U;Y_2|X_2,Q)]_+ \end{array}\right\} \quad (3)$$

where the function $[x]_+ = x$ if $x \geq 0$ and $[x]_+ = 0$ if $x < 0$. The auxiliary random variables $Q$ and $U$ are bounded in cardinality by $|\mathcal{Q}| \leq |\mathcal{X}_1| \cdot |\mathcal{X}_2| + 3$ and $|\mathcal{U}| \leq |\mathcal{X}_1|^2 \cdot |\mathcal{X}_2|^2 + 4|\mathcal{X}_1| \cdot |\mathcal{X}_2| + 3$, respectively.

*Remark 1:* The last bound in (3) indicates that there is a trade-off between the common rate and the secrecy level of confidential messages. As common rate $R_0$ increases, the secrecy level of confidential messages may get lower.

*Remark 2:* The rate-equivocation region (3) reduces to the capacity region of the MAC with degraded message sets given in [7] by setting $U = X_1$ and $R_e = 0$.

*Theorem 2:* The following region is an outer bound on the capacity-equivocation region $\mathscr{C}^I$ of the GMAC with one confidential message set:

$$\overline{\mathscr{R}}^I = \bigcup_{p(q,x_2)p(u|q)p(x_1|u)p(v|q)p(y,y_2|x_1,x_2)}$$
$$\left\{\begin{array}{l} (R_0, R_1, R_e): \\ R_0 \geq 0, \ R_1 \geq 0, \\ R_1 \leq I(U;Y|X_2,V), \\ R_0 + R_1 \leq I(U,X_2,Q;Y), \\ 0 \leq R_e \leq R_1, \\ R_e \leq I(U;Y|X_2,Q) - I(U;Y_2|X_2,Q), \\ R_0 + R_e \leq I(U,X_2,Q;Y) - I(U;Y_2|X_2,Q) \end{array}\right\}. \quad (4)$$

*Remark 3:* The last four bounds in the outer bound (4) match the last four bounds in the inner bound (3), and hence these four common bounds partially determine the boundary of the capacity-equivocation region $\mathscr{C}^I$.

We now study the case where perfect secrecy is achieved, i.e., user 2 does not get any information about confidential messages that user 1 sends to the destination. This happens if $R_e = R_1$.

*Definition 3:* The *secrecy capacity region* $\mathcal{C}_s^I$ is the region that includes all achievable rate pairs $(R_0, R_1)$ such that $R_e = R_1$, i.e.,

$$\mathcal{C}_s^I = \{(R_0, R_1) : (R_0, R_1, R_1) \in \mathscr{C}^I\}. \quad (5)$$

*Definition 4:* For a given rate $R_0$, the *secrecy capacity* $C_s^I(R_0)$ is the maximum achievable rate $R_1$ with confidential messages perfectly hidden from user 2, i.e.,

$$C_s^I(R_0) = \max_{(R_0,R_1)\in\mathcal{C}_s^I} R_1. \quad (6)$$

Although the outer bound (4) given in Theorem 2 provides only a partial converse to the capacity-equivocation region, it is sufficiently tight to serve as the converse to the secrecy capacity region and secrecy capacity (as a function of the common rate $R_0$).

*Theorem 3:* For the GMAC with one confidential message set, the following region is the secrecy capacity region:

$$\mathcal{C}_s^I = \bigcup_{p(q,x_2)p(u|q)p(x_1|u)p(y,y_2|x_1,x_2)}$$
$$\left\{\begin{array}{l} (R_0, R_1): \\ R_0 \geq 0, \ R_1 \geq 0, \\ R_1 \leq I(U;Y|X_2,Q) - I(U;Y_2|X_2,Q), \\ R_0 + R_1 \leq I(U,X_2,Q;Y) - I(U;Y_2|X_2,Q) \end{array}\right\}. \quad (7)$$

The secrecy capacity for a given rate $R_0$ is given by

$$C_s^I(R_0) = \max \min\{I(U;Y|X_2,Q) - I(U;Y_2|X_2,Q), \\ I(U,X_2,Q;Y) - I(U;Y_2|X_2,Q) - R_0\} \quad (8)$$

where the maximum is taken over all joint distributions $p(q,x_2)p(u|q)p(x_1|u)p(y,y_2|x_1,x_2)$. In both (7) and (8), the auxiliary random variables $Q$ and $U$ are bounded in cardinality by $|\mathcal{Q}| \leq |\mathcal{X}_1|\cdot|\mathcal{X}_2|+3$ and $|\mathcal{U}| \leq |\mathcal{X}_1|^2\cdot|\mathcal{X}_2|^2+4|\mathcal{X}_1|\cdot|\mathcal{X}_2|+3$, respectively.

*Remark 4:* If we let $R_0 = 0$ and $X_2 := \phi$, the GMAC with one confidential message set reduces to the case of a broadcast channel with confidential messages studied in [3] with the common rate being zero. For this channel, the secrecy capacity in (8) reduces to

$$\mathcal{C}_s^I = \max[I(U;Y) - I(U;Y_2)] \quad (9)$$

where the max is taken over all joint distribution $p(u,x_1)p(y,y_2|x_1)$. This is the same as the secrecy capacity given in Corollary 2 in [3].

## IV. DEGRADED GMAC WITH ONE CONFIDENTIAL MESSAGE SET

In this section, we study the degraded GMAC with one confidential message set, which generalizes the wire-tap channel studied in [2] to allow user 1 and user 2 (wire-tapper) to jointly send common information to the destination.

In the following, we define two classes of degraded GMACs with one confidential message set.

***Definition 5:*** The GMAC with one confidential message set is *physically degraded* if the transition probability distribution satisfies

$$p(y, y_2|x_1, x_2) = p(y|x_1, x_2)p(y_2|y, x_2). \quad (10)$$

***Definition 6:*** The GMAC with one confidential message set is *stochastically degraded* if its conditional marginal distribution is the same as that of a physically degraded GMAC, i.e., there exists a distribution $p(y_2|y, x_2)$ such that

$$p(y_2|x_1, x_2) = \sum_y p(y|x_1, x_2)p(y_2|y, x_2). \quad (11)$$

We note the following useful lemma.

***Lemma 1:*** The capacity-equivocation region of GMACs with confidential messages depends only on the marginal channel transition probability distributions $p(y|x_1, x_2)$, $p(y_1|x_1, x_2)$, and $p(y_2|x_1, x_2)$.

Based on Lemma 1, we have the following capacity-equivocation region for both physically and stochastically degraded GMACs with confidential messages.

***Theorem 4:*** For the degraded GMAC with one confidential message set, the capacity-equivocation region is given by

$$\mathscr{C}_d^I = \bigcup_{p(q,x_2)p(x_1|q)p(y|x_1,x_2)p(y_2|y,x_2)} \left\{ \begin{array}{l} (R_0, R_1, R_e): \\ R_0 \geq 0, R_1 \geq 0, \\ R_1 \leq I(X_1; Y|X_2, Q), \\ R_0 + R_1 \leq I(X_1, X_2; Y), \\ 0 \leq R_e \leq R_1, \\ R_e \leq I(X_1;Y|X_2,Q) - I(X_1;Y_2|X_2,Q), \\ R_0 + R_e \leq I(X_1, X_2; Y) - I(X_1; Y_2|X_2, Q) \end{array} \right\}, \quad (12)$$

where $Q$ is bounded in cardinality by $|\mathcal{Q}| \leq |\mathcal{X}_1| \cdot |\mathcal{X}_2| + 1$.

***Remark 5:*** The region $\mathscr{C}_d^I$ can be shown to be convex, and hence does not need further convexification.

The following secrecy capacity region and secrecy capacity (as a function of $R_0$) follow from Theorem 4.

***Corollary 1:*** For the degraded GMAC with one confidential message set, the secrecy capacity region is given by

$$\mathcal{C}_{s,d}^I = \bigcup_{p(q,x_2)p(x_1|q)p(y|x_1,x_2)p(y_2|y,x_2)} \left\{ \begin{array}{l} (R_0, R_1): \\ R_0 \geq 0, \ R_1 \geq 0, \\ R_1 \leq I(X_1; Y|X_2, Q) - I(X_1; Y_2|X_2, Q), \\ R_0 + R_1 \leq I(X_1, X_2; Y) - I(X_1; Y_2|X_2, Q) \end{array} \right\}. \quad (13)$$

The secrecy capacity as a function of $R_0$ is given by

$$C_{s,d}^I(R_0) = \max \min\{I(X_1; Y|X_2, Q) - I(X_1; Y_2|X_2, Q), \\ I(X_1, X_2; Y) - I(X_1; Y_2|X_2, Q) - R_0\} \quad (14)$$

where the maximum is taken over all joint distributions $p(q, x_2)p(x_1|q)p(y|x_1, x_2)p(y_2|y, x_2)$. In both (13) and (14), $Q$ is bounded in cardinality by $|\mathcal{Q}| \leq |\mathcal{X}_1| \cdot |\mathcal{X}_2| + 1$.

## V. GMAC WITH TWO CONFIDENTIAL MESSAGE SETS

In this section, we consider the general case of the GMAC with confidential messages, where the two users have common messages and each user has private (confidential) messages intended for the destination. For this case, the rate-equivocation tuple has five components and takes the form $(R_0, R_1, R_2, R_{1,e}, R_{2,e})$, where $R_{1,e}$ and $R_{2,e}$ are equivocation rates indicating the secrecy levels of confidential messages sent by user 1 and confidential messages sent by user 2, respectively.

***Theorem 5:*** The following convexified region of nonnegative rate-equivocation tuples is achievable for the GMAC with two confidential message sets:

$$\mathscr{R}^{II} = \textbf{Convex} \bigcup_{\substack{p(q)p(u|q)p(x_1|u)p(v|q)p(x_2|v) \\ p(y,y_1,y_2|x_1,x_2)}} \left\{ \begin{array}{l} (R_0, R_1, R_2, R_{1,e}, R_{2,e}): \\ R_0 \geq 0, R_1 \geq 0, R_2 \geq 0, \\ R_1 \leq I(U; Y|V, Q), \\ R_2 \leq I(V; Y|U, Q), \\ R_1 + R_2 \leq I(U, V; Y|Q), \\ R_0 + R_1 + R_2 \leq I(U, V, Q; Y), \\ (R_{1,e}, R_{2,e}) \in \mathcal{S}_e(R_0, R_1, R_2) \end{array} \right\}, \quad (15)$$

where

$$\mathcal{S}_e(R_0, R_1, R_2) = \bigcup_{\substack{(R_1', R_2'): (R_0, R_1', R_2') \in \mathcal{C}_{MAC}^p, \\ R_1 \leq R_1', R_2 \leq R_2'}} \left\{ \begin{array}{l} (R_0, R_{1,e}, R_{2,e}): \\ 0 \leq R_{1,e} \leq R_1 \\ R_{1,e} \leq [R_1' - I(U; Y_2|X_2, V, Q)]_+ \\ 0 \leq R_{2,e} \leq R_2 \\ R_{2,e} \leq [R_2' - I(V; Y_1|X_1, U, Q)]_+ \end{array} \right\} \quad (16)$$

where $\mathcal{C}_{MAC}^p$ is defined as

$$\mathcal{C}_{MAC}^p = \left\{ \begin{array}{l} (R_0, R_1, R_2): \\ R_0 \geq 0, R_1 \geq 0, R_2 \geq 0, \\ R_1 \leq I(U; Y|V, Q), \\ R_2 \leq I(V; Y|U, Q), \\ R_1 + R_2 \leq I(U, V; Y|Q), \\ R_0 + R_1 + R_2 \leq I(U, V, Q; Y) \end{array} \right\}. \quad (17)$$

The region $\mathcal{S}_e(R_0, R_1, R_2)$ can be characterized by an explicit form with inequality bounds only.

***Theorem 6:*** The set $\mathcal{S}_e(R_0, R_1, R_2)$ in (16) can be expressed in the following explicit form:

$$\mathcal{S}_e(R_0, R_1, R_2) \\ = \mathfrak{L}_1(R_0, R_1, R_2) \cup \mathfrak{L}_2(R_0, R_1, R_2) \cup \mathfrak{L}_3(R_0, R_1, R_2) \quad (18)$$

where

$$\mathfrak{L}_1(R_0, R_1, R_2)$$
$$= \left\{ \begin{array}{l} (R_{1,e}, R_{2,e}) : \\ 0 \leq R_{1,e} \leq R_1, \ 0 \leq R_{2,e} \leq R_2 \\ R_{1,e} \leq [I(U;Y|V,Q) - I(U;Y_2|X_2,V,Q)]_+, \\ R_{1,e} \leq [I(U,V;Y|Q) - R_2 - I(U;Y_2|X_2,V,Q)]_+, \\ R_{1,e} \leq [I(U,V,Q;Y) - R_0 - R_2 \\ \qquad - I(U;Y_2|X_2,V,Q)]_+ \\ R_{2,e} \leq [I(V;Y|U,Q) - I(V;Y_1|X_1,U,Q)]_+, \\ R_{2,e} \leq [I(U,V;Y|Q) - R_1 - I(V;Y_1|X_1,U,Q)]_+, \\ R_{2,e} \leq [I(U,V,Q;Y) - R_0 - R_1 \\ \qquad - I(V;Y_1|X_1,U,Q)]_+ \\ R_{1,e} + R_{2,e} \leq [I(U,V;Y|Q) - I(U;Y_2|X_2,V,Q) \\ \qquad - I(V;Y_1|X_1,U,Q)]_+, \\ R_{1,e} + R_{2,e} \leq [I(U,V,Q;Y) - R_0 \\ \qquad - I(U;Y_2|X_2,V,Q) - I(V;Y_1|X_1,U,Q)]_+ \end{array} \right\}$$

$$\mathfrak{L}_2(R_0, R_1, R_2)$$
$$= \left\{ \begin{array}{l} (R_{1,e}, R_{2,e}) : \\ 0 \leq R_{1,e} \leq R_1, \ R_{2,e} = 0 \\ R_{1,e} \leq [I(U;Y|V,Q) - I(U;Y_2|X_2,V,Q)]_+, \\ R_{1,e} \leq [I(U,V;Y|Q) - R_2 - I(U;Y_2|X_2,V,Q)]_+, \\ R_{1,e} \leq [I(U,V,Q;Y) - R_0 - R_2 \\ \qquad - I(U;Y_2|X_2,V,Q)]_+ \end{array} \right\}$$

$$\mathfrak{L}_3(R_0, R_1, R_2)$$
$$= \left\{ \begin{array}{l} (R_{1,e}, R_{2,e}) : \\ R_{1,e} = 0, \ 0 \leq R_{2,e} \leq R_2 \\ R_{2,e} \leq [I(V;Y|U,Q) - I(V;Y_1|X_1,U,Q)]_+, \\ R_{2,e} \leq [I(U,V;Y|Q) - R_1 - I(V;Y_1|X_1,U,Q)]_+, \\ R_{2,e} \leq [I(U,V,Q;Y) - R_0 - R_1 \\ \qquad - I(V;Y_1|X_1,U,Q)]_+ \end{array} \right\}$$

*Remark 6:* The region (15) reduces to the capacity region of the MAC [5] by removing the secrecy constraints ($R_{1,e} = 0, R_{2,e} = 0$) and setting $U = X_1$ and $V = X_2$.

*Remark 7:* The last two bounds in $\mathfrak{L}_1(R_0, R_1, R_2)$ indicate that there is a trade-off between the two equivocation rates $R_{1,e}$ and $R_{2,e}$, i.e., the secrecy levels achieved for the two confidential message sets $W_1$ and $W_2$.

We now study the case where confidential messages of each user are perfectly hidden from the other user. This happens when $R_{1,e} = R_1$ and $R_{2,e} = R_2$. The rate region that contains all these rate tuples is called the *secrecy capacity region* and is given by

$$\mathcal{C}_s^{II} = \{(R_0, R_1, R_2) : (R_0, R_1, R_2, R_1, R_2) \in \mathscr{C}^{II}\}. \quad (19)$$

The following inner bound on the secrecy capacity region $\mathcal{C}_s^{II}$ follows from Theorem 5 and Theorem 6.

*Corollary 2:* An inner bound on secrecy capacity region for the GMAC with two confidential message sets is given by:

$$\mathcal{R}_s^{II} = \textbf{Convex} \bigcup_{\substack{p(q)p(u|q)p(x_1|u) \\ p(v|q)p(x_2|v) \\ p(y,y_1,y_2|x_1,x_2)}} \left\{ \mathcal{R}_{s,1} \cup \mathcal{R}_{s,2} \cup \mathcal{R}_{s,3} \right\}$$

where

$$\mathcal{R}_{s,1} = \left\{ \begin{array}{l} (R_0, R_1, R_2) : \\ R_0 \geq 0, R_1 \geq 0, R_2 \geq 0, \\ R_1 \leq I(U;Y|V,Q) - I(U;Y_2|V,X_2,Q), \\ R_2 \leq I(V;Y|U,Q) - I(V;Y_1|U,X_1,Q), \\ R_1 + R_2 \leq I(U,V;Y|Q) - I(U;Y_2|V,X_2,Q) \\ \qquad - I(V;Y_1|U,X_1,Q), \\ R_0 + R_1 + R_2 \leq I(U,V,Q;Y) \\ \qquad - I(U;Y_2|V,X_2,Q) - I(V;Y_1|U,X_1,Q) \end{array} \right\}$$

$$\mathcal{R}_{s,2} = \left\{ \begin{array}{l} (R_0, R_1, R_2) : \\ R_0 \geq 0, R_1 \geq 0, R_2 = 0, \\ R_1 \leq I(U;Y|V,Q) - I(U;Y_2|V,X_2,Q), \\ R_0 + R_1 \leq I(U,V,Q;Y) - I(U;Y_2|V,X_2,Q) \end{array} \right\}$$

$$\mathcal{R}_{s,3} = \left\{ \begin{array}{l} (R_0, R_1, R_2) : \\ R_0 \geq 0, R_1 = 0, R_2 \geq 0, \\ R_2 \leq I(V;Y|U,Q) - I(V;Y_1|U,X_1,Q), \\ R_0 + R_2 \leq I(U,V,Q;Y) - I(V;Y_1|U,X_1,Q) \end{array} \right\}$$

## VI. CONCLUSIONS

We have studied the capacity-equivocation region of the GMAC with confidential messages. For the GMAC with one confidential message set, we have derived inner and outer bounds on the capacity-equivocation region. Although the two bounds only match partially, they are tight enough to characterize the secrecy capacity region, where confidential messages sent by user 1 are perfectly hidden from user 2. For the degraded GMAC with one confidential message set, we have established the capacity-equivocation region. In [6], we have also studied two example GMACs with one confidential message set, the binary degraded GMAC and the Gaussian GMAC.

We have further obtained an achievable rate-equivocation region (inner bound on the capacity-equivocation region) for the general case of the GMAC with two confidential message sets. We have shown that the achievable rate-equivocation region for the case of two confidential message sets carries a new feature of a trade-off between the two equivocation rates corresponding to the two confidential message sets sent by the two users.